# SoK: A Study of the Security on Voice Processing Systems

Robert Chang, Logan Kuo, Arthur Liu, and Nader Sehatbakhsh
S²ysArch Lab, UCLA

*Abstract*—As the use of Voice Processing Systems (VPS) continues to become more prevalent in our daily lives through the increased reliance on applications such as commercial voice recognition devices as well as major text-to-speech software, the attacks on these systems are increasingly complex, varied, and constantly evolving. With the use cases for VPS rapidly growing into new spaces and purposes, the potential consequences regarding privacy are increasingly more dangerous. In addition, the growing number and increased practicality of over-the-air attacks have made system failures much more probable. In this paper, we will identify and classify an arrangement of unique attacks on voice processing systems. Over the years research has been moving from specialized, untargeted attacks that result in the malfunction of systems and the denial of services to more general, targeted attacks that can force an outcome controlled by an adversary. The current and most frequently used machine learning systems and deep neural networks, which are at the core of modern voice processing systems, were built with a focus on performance and scalability rather than security. Therefore, it is critical for us to reassess the developing voice processing landscape and to identify the state of current attacks and defenses so that we may suggest future developments and theoretical improvements.

## I. Introduction

5G is around the corner and the Internet of Things is becoming a reality. Voice assistants like Siri, Alexa, and Echo can now turn on your lights and answer your questions in a blink of an eye. The assistants are beginning to understand you with higher accuracy and might just stop telling you to repeat yourself. This is all propelled by the advancement of Voice Processing Systems (VPS). However, like any new technology, it might be wise to assess just how safe it is before deciding whether to fully embrace the convenience of executing every command through your voice.

Technologies around Voice Processing Systems rapidly improves in terms of its practicality and accuracy, society has gravitated towards smart devices that implement convenient features which assist people's daily life. VPS aims to classify, describe, or generate audio from input audio samples which requires the utilization of machine learning. Many intelligent devices created from these machine learning datasets provide features such as voice control and voice recognition that deeply influence our daily life. Some examples include voice recognition passcodes to unlock private environments as well as commanding the intelligent device to make an action for us (e.g., making a purchase or playing said music). Voice processing systems encourage users to initiate voice commands as a quality-of-life improvement towards their lives. As we realize the impressiveness and increasing popularity VPSes have due to its lasting impacts and how it can deeply change our lives for the better, the rapid growth of this field has also created red flags in terms of adversarial attacks and privacy concerns.

With the recent focus on methods to break device security and/or privacy ranging from password cracking and malware installation to physical side-channels and leakage, VPSes have presented themselves as an interesting new area of focus. Unlike the targeting of direct defense mechanisms such as multi-factor authentication and application permissions which were constructed with a focus geared towards security and therefore have effective safeguards in place to slow down attackers, VPSes were designed with a focus on usability and simplicity for its users and therefore do not offer much resistance against new cyber-attacks. In 2017, hidden voice attacks were used to successfully break certain white-box systems by injecting voice commands that were properly processed and understood by VPS machine learning models while remaining inaudible to humans [10]. Since then, new attacks and improvements have been developed which instead target the entirety of the signal processing phase of VPSes thus generalizing the attack from specific white-box models to black-box systems [6, 8, 13, 14, 18, 19, 20, 21].

As recent works have built upon these developed attacks by increasing their impact range and rate of accuracy, it has become more imperative to better understand the privacy dangers that lie in the use of various VPS technologies [4, 9, 11, 14, 17].

In this work, we present a systematic analysis on the current threats on VPS and study the proposed defense mechanisms against those attacks.

Specifically, this paper makes the following contributions:

- **A systematic analysis of attack strategies on VPS:**

    To fully understand the existing threats on VPS, we break down the threat model into *five* different categories and analyze how each category can impact the overall security and/or privacy of the overall system.

- **A taxonomy on proposed defense mechanisms for VPS:**

  By providing a thorough analysis on the state-of-the-art, we systematically categorize the existing defense mechanisms for securing VPS.

- **Providing insights on future attacks and defense trends in VPS:**

  Using our observations, we share our insights on how the future would look like for VPS security?

This paper is organized such that all relevant information needed to understand the general trends in both the attacks and defenses of voice processing systems is contained in Section II. Furthermore, Section II will highlight the major use cases of voice processing and demonstrate the input flow for a given sound up to the machine learning algorithm stage. In addition, supplemental information will be provided on general machine learning processing techniques and algorithms that are often used by voice processing systems. Sections III and IV will respectively traverse and explain the categories that we used to group related patterns observed in the voice system landscape. Section V will then take the assigned groupings from Sections III and IV and provide additional commentary and new suggestions for future areas of focus and critical sections for improvement. Conclusion is presented in Section VI.

## II. BACKGROUND

### A. Primary Subsets of Voice Processing Systems (VPS)

VPS can be used for two main scenarios, one of them being *Automatic Speech Recognition* and the other being *Automatic Speaker Identification*.

*1) Automatic Speech Recognition (ASR)*: The purpose of ASR is to achieve speech-to-text with a high level of accuracy. The system takes in audio input and translates them into a text output for a certain language. ASR is an important aspect of Internet of Things (IoT) systems in which voice assistants, such as Apple Siri, Google Alexa, Amazon Echo, etc., are at the center of controlling smart devices. ASR is also commonly used in captioning of live broadcasts on the news, during conferences, and in virtual meetings. Some popular productized ASR services include Google Cloud Speech-to-Text, Dragon Naturally Speaking, Amazon Transcribe, and Microsoft Azure Speech to Text. Many Automatic Speech Recognition (ASR) services require some utilization of VPS to allow users to fully interact with smart devices.

*2) Automatic Speaker Identification (ASI)*: The purpose of ASI is to be able to match the audio source to its speaker. In addition, ASI are often extended by automatic speaker verification (ASV) systems to decide if an identity claim on an audio source is true or false. ASI is used as biometric recognition and can be placed in the security system similar to other biometrics such as fingerprint and iris scanning.

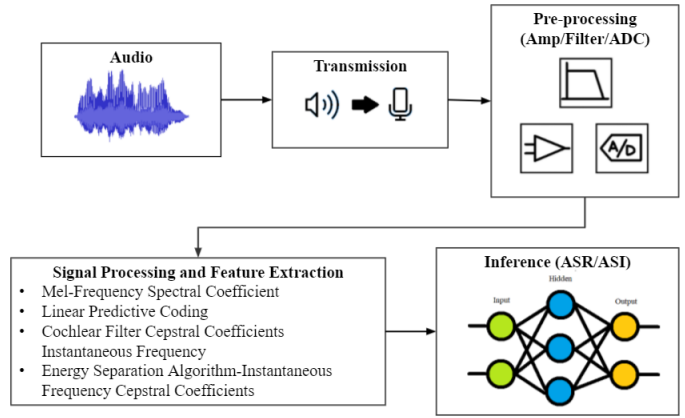

Figure 1: Pipeline diagram demonstrating the VPS functional process. It includes the Audio and Transmission (which we grouped together as Audio Input), Pre-processing, Signal Processing/Feature Extraction, and Machine Learning Inference stages.

### B. Primary Functional Blocks of VPS

Modern VPS are split into four functioning blocks to convert the audio input to the desired output, whether it be speech to text or speaker identification.

The first block is the microphone, and it converts audio to electric signals. The front-end applies pre-processing functions such as noise filtering, sampling, and digitizing. The main signal processing blocks are designed based on the application of the VPS. The most popular ones function in the frequency domain and apply the Fast Fourier transform algorithm to form spectrograms. The spectrogram typically contains information about the frequency, timescale, acoustic frame, and the energy used to produce the sound and is injected to classifiers for inference. In the following, we describe these four blocks in more detail.

*1) Audio Input:* The first part of the system is the audio input. Raw audio inputs are injected into the voice processing system either through a microphone or directly into the audio input system. The microphone serves as a tool to convert sound from an analog environment to a digital form. Injecting audio directly into the system makes the input replayable and has less volatility from the environment. Audio input through the microphone will have its quality impacted by the range of the audio source, noise in the background, echo from the surrounding, and sensitivity of the microphone hardware.

If adversaries are able to obtain the raw audio input that is supposed to be run through the voice processing system, then they can directly manipulate the input of the system.

This could be categorized as *advarsarial attacks* include attacks such as *feature injection* or even *poisoning data*. By obtaining information about the input and looking at what the system would output, adversaries are able to make the connection of the processes made onto the input signal that would give said output. This would give the attackers a lot of free room to explore even more vulnerabilities in the system by being able to access the inputs and feeding many attacks and modified raw audio inputs for information gain or malicious activity.

The front end is also vulnerable to audio jamming by adding audio features undetectable by the human ear. The jamming signal significantly raises the noise floor to reduce signal to noise ratio requirements for proper signal processing.

*2) Audio Pre-processing*: The second part of the system is the audio pre-processing stage which aims to extract the important audio information from the raw input. The process involves the removal of unwanted partitions such as any background noise that aims to lower the resolution quality of the audio input. The result would be a cleaner audio file without major signal changes. The block utilizes analog filtering and audio signal amplification before relaying the signal processing. In addition, noise reduction, harmonic enhancement, and other processes are used to remove unnecessary information from the audio input.

By understanding what parts of the audio signals are deleted or removed, attackers can avoid detection through understanding what is processed or filtered out of the signal before entering the next block of processing data. On the flip side, adversaries are also able to sabotage accurate or important partitions of the audio input if the pre-processing block of the system is compromised. Information can be stolen and even corrupted if adversaries are able to control the algorithms that involve the deletion and filtering of the "unwanted" parts.

*3) Signal Processing*: The third part of the system is the signal processing. This block prepares the time domain signal into features that are ready to be analyzed and inference by machine learning algorithm. The best signal processing aims to retain and best approximate audio features that human ears can capture.

Some of the most common processing algorithms include mel frequency cepstrum coefficient (MFCC). There are also many other techniques including Mel-Frequency Spectral Coefficients (MFSC), Linear Predictive Coding, Perceptual Linear Prediction, Constant-Q Cepstral Coefficients, and Cochlear Filter Cepstral Coefficients to extract various aspects of the audio features.

This block is also the second most flexible and diverse portion next to the machine learning algorithm. Defense relies on this stage to insert detection mechanisms that attackers try to evade.

*4) Machine Learning Inference*: The final part of the system is machine learning inference. The machine learning (ML) algorithm depends on the application of the voice processing system. For ASR the ML aims to use the audio feature and correctly assign word labels for the audio input. Popular algorithms often implement a convolution neural network trained with supervised learning as the inference system. ASI system uses this stage to match the input against a feature of the database of speakers to infer the identity. An alternative approach is the use of a Hidden Markov Model (HMM) which is a statistical model that is structured with numerous layers. Although the number of layers can vary, HMMs typically have at least three layers with their own unique goals. The first is focused on the acoustic level and its job is to determine if the guessed phoneme, the smallest unit of distinct sound, is actually the correct phoneme that was heard. The second layer then is responsible for checking if multiple phonemes are probabilistically likely to be placed next to one another in an actual word. The third layer then operates on a word level in order to verify if multiple predicted words from the second layer actually make sense to be placed together in a sentence (both logically and grammatically). If at any point, a layer determines that it made a mistake with the previous predictions, it will backtrack to the previous layer in order to select a different choice. Both HMMs and neural networks have their pros and cons and so in some cases, *hybrid* approaches are adopted which utilize both kinds of models in order to make their predictions.

There is a diverse number of ML algorithms that target different features from the audio signal. Some target temporal dependency [1, 7, 11, 12], high order frequency behavior [1, 6, 10, 13, 18], baseband aliases [1, 14], sound perturbation [2, 4, 5, 9, 17, 18], mechanical vibrations [8], etc. Additionally, ML may also extract different parameters from these features including duration and power signatures.

Attackers have since moved from requiring white-box information to attempts in creating algorithms that are effective with black box information. In the next section, we provide our detailed analysis on the existing attack vectors on a VPS.

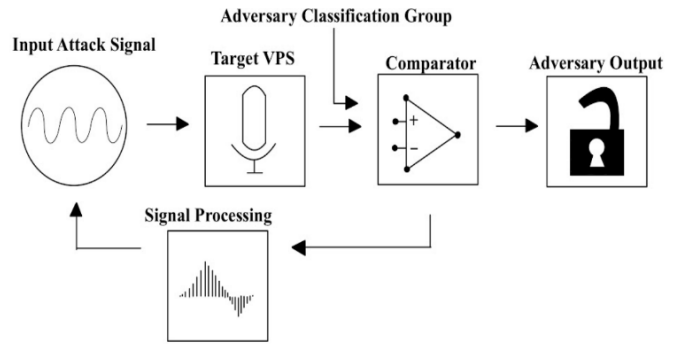

Figure 2: Iterative attack model consisting of a feedback loop. If the output of the target VPS is not the one wanted by the adversary classification group, we reprocess the audio signal and attack the VPS again.

III. ATTACK MODEL

Attacks on VPS are best understood in relation to the context of the victim system, what exactly is being done, where the vulnerabilities are. To break down the details of each attack for fair analysis it is important to define categories to be used in comparisons. Specifically, we define five different categories that can distinguish the existing attacks. In the following, we describe them in more detail.

*A. Consequences*

The consequence of a successful attack is dependent on what the original system was meant to perform versus what the compromised system actually does. There are a few main types of attacks that we will call denial of service, targeted control of service, and leakage of information from the service.

*1) Denial of Service*: Denial of service can happen in many forms, but all symptoms of the attack culminate in the scenario where a legitimate user is attempting to use the VPS, but the result is incorrect. For example, the input is not captured due to jamming, therefore the output simply could not output the correct inference [6, 21].

*2) Targeted Control of Service*: Targeted control of service is the scenario where the attacker is able to manipulate the input to the VPS and obtain a targeted output from the system. This can happen with or without the presence of legitimate user commands. A popular attack is known as hidden voice attack, which broadcasts sound incomprehensible to humans but recognized as proper command by VPS [6, 8, 10, 13, 14, 16, 18, 19, 20].

*3) Leakage of Information*: Leakage of information is best described and dependent on the implementation of the VPS. The purpose of this attack is to collect voice information that is being processed by the system and to fingerprint the usage data and behavior of the legitimate user. This type of attack is most feasible when the model inference needs to happen in the cloud and audio information may be intercepted or reconstructed with side channel attacks at any point in the VPS [3].

*B. Domain*

Attacks can vary based on their purpose and malicious intent as seen from Section III-A, whether it be the denial of service, system control override, or leakage of information. Due to the large number of varying attacks with different purposes, there exists many ranges and domains each attack traverses. We will classify the traversal domains into two different main groups.

*1) Physical Attacks*: When attackers use the physical type of attacks, the attack from the physical audio source is first transmitted through an open channel through the environmental air, which is also known as an over-the-air attack, before reaching to the device side. Because the attack is transmitted through the air into the device, environmental factors in the surrounding perimeter such as background noise may affect the transmission of the audio signal. For physical attacks, it is important to note the distance the audio signal is traveling. This is the case because as the distance during the transmission of the signal increases, the strength of the signal decreases. This imposes a constraint of the maximum distance that a signal can be read on a device to obtain enough energy from that signal. Another important factor to consider is the environment surrounding itself. The input signal would be reflected across many surfaces which results in the capture signal on the device to be a composition of signals instead of the original raw signal. Because of this mixed signal composition, adversaries can exploit this fact and broadcast their attacks while hiding it with the human voice by overlaying the signals. This attack would compromise VPS, especially systems with ASR functionalities since they would misinterpret voice commands and function as the attacker has intended. Therefore, there are many factors to be considered for physical attacks done over-the-air against voice processing systems in terms of the physical space and environment.

*2) Digital Attacks*: The digital attacks can start as early as the pre-processing stage where digital attacks can be inputted in the module. If attackers can figure out signal properties and patterns as well as the desired commands to use on an attack, then they will be able to determine a raw audio signal that preserves the features and commands which is incomprehensible to the human ear. This process is usually done through decoding parts of the signal which comes from the pre-processing module extracting some amount of loss in the signal. From these losses that the pre-processing module has extracted, attackers can correspond these losses to the actual acoustic feature and thus get an idea of the architecture and structure of the voice processing module. Other strategies of digital attacks also include knowing the voice activity detection threshold value which determines whether an audio segment is considered a command or not. If attackers can find the voice activity detection value, they will be able to optimize their attacks to compromise the VPS. Many other digital attack models can be explored since signal processing itself is very broad and has a high chance that attacks can go through this module block.

*C. Generality*

Generality defines how well the attack applies to or could be applied to various VPSes and system configurations.

*1) Universal*: The attack is evident to work in both ASR and ASI systems. The type of attack is also likely extendable to all commercial datasets and machine learning models.

*2) Specific*: The attack is targeted at a particular group of individual VPSes and is likely not effective against systems outside of the targeted group.

*D. Knowledge of the Victim System*

Knowledge of the victim system describes the amount of information about the architecture and VPS implementation that the attacker needs to implement his/her attack.

*1) Black Box*: The attacker does not need information about the system beyond knowing if the system is a voice processing system.

*2) White Box*: The attacker requires the information of the system components, setup, ML algorithm, and potentially the model parameters as well.

*E. Attack Mechanism*

Attack mechanism describes the actual implementation of each attack and what area of vulnerability the attack is exploiting.

TABLE I
Attack Model Labeling

| Attack | Consequence | Domain | Generality | Knowledge | Mechanism |
|---|---|---|---|---|---|
| [1] 2020 | Targeted Control of Service | Digital | Specific | Black Box | Speaker Impersonation |
| [2] 2019 | Denial of Service | Digital | Specific | Black Box | Synthetic Speech |
| [3] 2020 | Information Leakage | Physical | Specific | Black Box | VPS Side-Channel |
| [4] 2020 | Denial of Service | Physical | Specific | White Box | Hidden Voice Command |
| [5] 2018 | Denial of Service | Digital | Specific | White Box | Synthetic Speech |
| [6] 2017 | Targeted Control of Service, Denial of Service | Physical | Universal | Black Box | Hidden Voice Command |
| [8] 2019 | Information Leakage, Targeted Control of Service | Physical | Universal | Black Box | Hidden Voice Command |
| [9] 2020 | Targeted Control of Service | Physical, Digital | Universal | Black Box | Synthetic Speech |
| [10] 2017 | Targeted Control of Service, Denial of Service | Physical | Universal | White Box | Hidden Voice Command |
| [11] 2019 | Denial of Service | Digital | Universal | Black Box | Synthetic Speech |
| [13] 2017 | Targeted Control of Service, Denial of Service | Physical | Universal | Black Box | Hidden Voice Command |
| [14] 2018 | Targeted Control of Service | Physical | Universal | Black Box | Hidden Voice Command |
| [16] 2019 | Targeted Control of Service | Physical | Specific | White Box | Hidden Voice Command |
| [17] 2020 | Targeted Control of Service | Physical, Digital | Specific | White Box | Synthetic Speech |
| [18] 2019 | Targeted Control of Service | Physical, Digital | Universal | Black Box | Hidden Voice Command |
| [19] 2020 | Denial of Service, Targeted Control of Service | Digital | Universal | Black Box, White Box | Hidden Voice Command, Synthetic Speech |
| [20] 2021 | Targeted Control of Service, Denial of Service | Physical, Digital | Universal, Specific | Black Box, White Box | Hidden Voice Command, Synthetic Speech |
| [21] 2019 | Denial of Service | Physical | Universal | Black Box | Hidden Voice Command |

*1) Hidden Voice Command:* Hidden Voice Command is a family of attack mechanisms against VPS that exploits sounds not recognized by humans but recognized by VPS. This is achieved through various means including Time Domain Inversion [18], Random Phase generation [18], High Frequency (ultrasonic) [6, 10, 13, 18], Time Scaling [18], etc. Hidden voice commands are usually manufactured from scratch without any base sample voice input. This typically targets the signal pre-processing and processing stages.

*2) Synthetic Speech*: Synthetic Speech also known as adversarial sample is a family of attack mechanisms that tries to generate modified audio samples from valid voice command input samples that would fool the VPS or transfer its control to the attacker. A common theme includes digitally manipulating input samples and then digitally transmitting it so that the VPS will create errors and output different results. This typically attacks the machine learning stage.

*3) Speaker Impersonation*: Speaker impersonation is a family of attack mechanisms that tries to fool ASI systems through mimicking audio features present in the correct owner. There are a few types of attack. The first is a human impersonation by a malicious attacker pretending to sound like a legitimate speaker. The second is synthetic speech that generates voice samples that mimic the original speaker. The third is voice conversion that records an attacker's audio sample and attempts to convert it with features similar to the victim's voice. The fourth type is a replay attack that records audio samples from the victim and plays it against the ASI system.

*4) VPS Side-Channel*: VPS Side-Channel is a family of attack mechanisms that ultimately try to obtain information that may be sensitive. The leakage is based on the implementation of a computer system. By being able to observe and gather data using side-channels, we can combine our intel with unhidden voice attacks. This can be done by exploiting the gained information and compromising the system security. For example, voice data leaked from side-channels can be used to redirect access control of a system to an adversary's node, enabling malicious unauthorized commands from the attacker.

## IV. DEFENSE MODEL

Defense models are usually generated from previously created adversarial attacks on certain open-source data sets, meaning that it is specialized in preventing attacks from a type of known attack. This means that there is no single defense system that can withstand every known attack based on the one particular model and framework. This part of the paper aims to gather certain specialized defense mechanisms and provide its target capabilities against certain adversarial attacks.

### A. Mitigated Attacks

An important trait for analyzing and comparing different defense models is the type of attack mechanisms that the defense is able to mitigate or was engineered to counteract. Such attack mechanisms for emerging VPS attacks can vary greatly and therefore, we will be applying the same attack mechanism categories that we used previously in the paper. For definitions and explanations of the following labels, see Section III-E.

1) Hidden Voice Command
2) Synthetic Speech
3) Speaker Impersonation
4) VPS Side-Channel

### B. Domain of Defense Implementation

Since attacks can be very situational and instantaneous whether it be physical attacks transmitted over the air or digital attacks through signal processing, there are many different implementations of defense mechanisms. The paper will discuss (1) hardware implementation and (2) software implementation defense mechanisms that are prevalent in protecting against certain attacks.

*1) Hardware Implementation*: These implementations tend to be more secure and robust against adversarial attacks since there are hardware components in devices that are dedicated to security consisting of many attack prevention mechanisms. Hardware implementations are also almost more efficient and faster performing. However, hardware implemented defense mechanisms would be more financially costly, resource impactful, and time demanding due to its tangibility and specialized functionality. Some examples of hardware implementation defenses include microphones with the capability to recognize processed malicious signals or even microphone jamming hardware that prevents audio input when voice commands are not wanted in certain situations [10, 21].

*2) Software Implementation*: The majority of defense mechanisms are focused on the software implementation of defense mechanisms. Software implemented defense mechanisms have some universality towards them since there are only a handful of ways to exploit the software through software. Software-based defenses are usually slower performing than hardware-based ones but faster to develop. Software defense mechanisms tend to focus more towards increasing the robustness of machine learning models, as well as introducing algorithms that can detect differences and anomalies among data results. An example of software implemented defense mechanisms include pre-training an observed neural network with datasets inclusive of adversarial attacks as well as random clean samples to have the neural network be wary of some adversarial attack, just like how a vaccine works. Software mechanisms, as we can see, are focused on pattern recognition and attack identification with algorithms and programs preset before system operation [1, 7, 10, 12, 14, 15].

### C. Generality

This section will classify defense mechanisms based on how wide of a variety of platforms that they apply to or can be potentially applied to. These classifications were based on experimental results of certain adversarial attacks.

*1) Universal*: These defense mechanisms seem to be able to be implemented with a common factor of products, meaning that the defense mechanisms can either protect many differing datasets with a universal hardware component or a universal algorithm or machine learning model.

*2) Specific*: Specific defense mechanisms include results on certain datasets and have not yet had the results be replicated through other datasets. More tests are needed in order to increase either the generality of the mechanism or the practicality of the defense mechanism implementation.

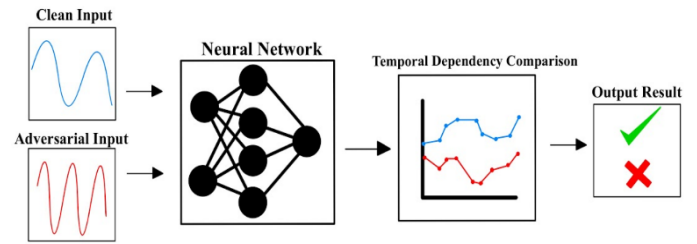

Figure 3: The architecture analyzes the clean input and adversarial input through a neural network and compares its temporal dependency to find if there are discrepancies.

TABLE II
Defense Model Labeling

| Defense | Mitigated Attacks | Domain | Generality | Knowledge | Mechanism |
|---|---|---|---|---|---|
| [1] 2020 | Synthetic Speech, Hidden Voice Command, Speaker Impersonation | Software | Universal | Detectable ML Misbehavior, Detectable Frequency Corruption | Anomaly Pattern Detection |
| [7] 2019 | Synthetic Speech, Hidden Voice Command, Speaker Impersonation | Software | Universal | Detectable ML Misbehavior – Temporal Dependency | Anomaly Pattern Detection – ML Classification, Input Transformation |
| [10] 2017 | Hidden Voice Command | Hardware | Universal | Detectable Frequency Corruption – High Frequency | Anomaly Pattern Detection – Statistical Comparison, Input Transformation |
| [10] 2017 | Hidden Voice Command | Hardware | Specific | Detectable Frequency Corruption – High Frequency | Input Transformation |
| [10] 2017 | Hidden Voice Command | Software | Universal | Detectable Frequency Corruption – High Frequency | Anomaly Pattern Detection – ML Classification |
| [12] 2020 | Synthetic Speech | Software | Universal | Detectable ML Misbehavior – ML Activation | Anomaly Pattern Detection – Statistical Comparison |
| [14] 2018 | Hidden Voice Command | Software | Universal | Detectable Frequency Corruption – Low Frequency | Anomaly Pattern Detection – Statistical Comparison, Input Transformation |
| [15] 2019 | Synthetic Speech, Hidden Voice Command, Speaker Impersonation | Software | Universal | Detectable ML Misbehavior – ML Activation | Anomaly Pattern Detection – ML Classification |

*D. Knowledge of Attack System*

This category describes what attack properties the defense mechanism tries to target in order to detect VPS attacks.

*1) Detectable ML Misbehavior*: Attacks will cause detectable ML behavior patterns. The data can often be parsed in a specific manner to expose differences in internal dependencies between the unaltered and modified input samples.

  *a)* ML activation is systematically different between adversarial and benign inputs.

  *b)* Temporal dependency is systematically different between adversarial and benign inputs.

*2) Detectable Frequency Corruption:* Certain frequency ranges contain detectable malicious features that are otherwise not present or altered in normal audio samples. Frequency corruption often arises because of purposely injected noise or frequency shifts.

  *a)* High frequency contents contain detectable malicious features.

  *b)* Low frequency contents contain detectable malicious features.

*E. Defense Mechanism*

Defense mechanism describes the actual implementation of each defense and the method by which the defense response operates.

*1) Anomaly Pattern Detection*:

  *a) ML Classification:* The defense utilizes machine learning algorithms to detect anomalous patterns. The defense can modify the existing VPS by adding new layers or subsystems to detect attacks. Some defenses also choose to implement a separate classifier that monitors the status of the VPS such as node activations.

  *b) Statistical Comparison:* Defenses monitor certain status of the VPS during audio processing and generate a statistical

distribution. At runtime, the defense compares the VPS behavior from the current input with the expected behavior from prebuilt distribution. The properties monitor often comes in the form of frequency spectrums or ML activations. In terms of frequency spectrums, defense typically analyzes power or frequency patterns that are outside audio ranges.

*2) Input Transformation*: This technique was first explored in the image domain since input transformation was a common method to prevent image adversarial attacks. The main strategy is to remove the adversarial and hidden contents of audio samples. A few techniques include quantization, downsampling, autoencoding, local smoothing, and frequency noise canceling. All of these techniques aim to remove noise from the audio sample and only collect the semantic portion of the input. These methods assume that attacks will thus be removed directly and omit the need for detection.

V. DISCUSSION

*A. Trends of Attack*

We have classified and created categories of the main consequences of recent attacks into denial of service attacks, targeted control of service, and information leakage attacks.

*1) Denial of Service Attacks*: Firstly, we have seen early papers describe attacks that aim to deny the service of some audio device or machine learning network. Denial of service attacks seem to be more easily conducted than other attacks since the goal of the attack is to return an error in order to deny the service of the observed device. The error that the attack aims to return can be regarded as a garbage value, which is sufficient for attacks that only require the denial of service.

*2) Targeted Control of Service Attacks*: On the other hand, attacks that are targeted control of service require much more precision. Instead of only returning a dummy value to deny the service, targeted control of service attacks need to find some way to gain exclusive control, meaning some transfer of control process needs to occur in the victim system device. The attacks that result in sensitive information leaks enable targeted control of service attacks. Since targeted control of service attacks requires a certain way to gain access or control of the victim system, there needs to be some prior knowledge these adversarial attacks must gain first in order to succeed effectively.

*3) Information Leakage Attacks*: Information leakage attacks are the first step of more complicated and complex attacks since knowing more about the victim system will result in better understanding in the defenses adversarial attacks must avoid and bypass. In order for attacks to be much more realistic and practical, these attacks also should be hidden from normal device users, which add another layer of complexity for any type of adversarial attack.

*B. Practicality of Attacks*

The practicality of attacks have also been improving over the years. Many of the early attacks aim to demonstrate proof of concepts on vulnerability exploits. As research in the types of attacks mature, researchers have begun to expand the capabilities of such attacks. For example, early synthetic speech attacks have been requiring white box knowledge [5, 10], but recently more attacks have been conducted on a black box setting [2, 9]. In addition early attacks abstracted away from the real world setting by digitally injecting attacks directly to the VPS [2, 5, 10]. In contrast, more recent papers [9, 17] have demonstrated successful synthetic speech attacks on VPS over the air. Similar trends can be seen in research in hidden voice attacks, speaker impersonation attacks, and VPS side channel attacks.

*C. Trends of Defense*

As we have discussed and identified trending attacks, we also aim to discuss some defense mechanisms that are currently being developed. We see trends in defending against synthetic speech adversarial attacks through *pattern recognition and anomaly detection*. What we have concluded is that no matter how small a feature injection from an attack, or how nuance a signal is changed to form an attack, there is still a discrepancy to be found inside the original audio signal. Adversarial attacks and defense protocols are always striving to compete against each other with adversarial attacks always having the first move due to its freedom in what type of system to attack and the methodology.

This is also true in the fact that security and defense mechanisms have always been a second priority when developing a system or device. Some specific defense mechanisms we have gathered include input transformation techniques used to observe temporal dependency, as well as pre-training machine learning systems to account for adversarial attacks in order to detect extreme anomalies.

As we have discussed above, there needs to be a sufficient amount of currently existing adversarial attacks in order to properly train a machine learning system to counter these attacks through anomaly detection. Input classification methods may be more costly or expensive since they would transform all audio signal data and then rely on observing the temporal dependency on its data set in order to come up with a consensus of attacks and clean audio input. These methods we have outlined focus towards pattern recognition and always compare an adversarial attack using some sort of metric in order to differentiate the two. Some future defense mechanisms that we encourage can be towards creating embedded systems, since these types of defense mechanisms can be more secure. Even though having hardware implementations like an embedded chip can be more costly, it may be able to prevent some types of attacks by having a layer of security adversarial attacks are not able to reach.

*D. Practicality of Defenses*

As of now, a single universal defense mechanism that can prevent a large general field of attack, such as audio attacks, does not seem to be possible since adversarial attacks in a certain field can use many different mechanisms and techniques as we have discussed above. We can now see that the reason why defenses always come after attacks is because adversarial attacks need to be prevalent in order for defense mechanisms to prevent these specific attacks. We try to generalize and group defense mechanisms which is still a work in progress. In addition, there is a lack of evidence regarding how easily the defense mechanisms can be deployed on commercial systems since the authors of defense papers often implement a custom system to verify performance.

*E. Comparison to Other Machine Learning Systems*

As mentioned throughout this paper, machine learning, especially neural networks, are a key part of many existing VPSs. While ML generally provides a significant improvement for these speech recognition systems, it also makes them vulnerable to the existing threats on ML systems, collectively referred to as *adversarial machine learning*. The attacks in this domain are typically categorized into *(1) Poisoning Attacks*: It occurs when the adversary is able to inject malicious data into the model's training pool, causing it to learn something it shouldn't have [22-25]. *(2) Evasion Attacks*: It happens when the adversary, by crafting noisy samples (aka adversarial samples), *fools* the ML model to incorrectly label the given input [25-29]. *(3) Privacy Attacks*: Where the adversary steals some sensitive information for the model. This information could be extracting the model [30-32] or inverting it [33], input data [35], or some aspects of it [34].

Compared to the other applications for machine learning, such as image recognition, vision, etc., to the best of our knowledge, there is no known privacy and/or poisoning attacks on VPSs. However, *we believe this a very possible research direction in future*, as these VPSs share a similar threat model with other ML models such as vision. Possible attack scenarios include reverse-engineering the VPS in order to extract its ML model, poisoning data sets (e.g., voice command samples) to create backdoors, etc.

In contrary, as described in this paper, there are a number of evasion attacks for VPS. Most of the literature on evasion attacks though, has been focused on the computer vision. In that context, adversarial samples arise from introducing perturbations to an image that do not alter its semantics (as determined by human observers), but that lead machine learning models to misclassify the image.

In the context of audio, however, evasion attacks can happen by manipulating the input sound by adding perturbations (e.g., by adding inaudible signals, silence, noise, etc.) some of which are also common in other ML systems (e.g., adding noise), but some other are unique to VPS (e.g., inaudible signals).

Besides having different methods for creating adversarial samples, successful evasion attacks are typically more difficult to carry out on VPSs. In evasion attacks, although sending adversarial samples to a VPS is relatively simple (since they are typically more accessible than other ML systems), conducting successful attacks is much more challenging because these ML systems normally rely on signal processing stage extracted features as inputs rather than raw image data.

VI. CONCLUSION

In this paper we have systematically analyzed state of the art attacks against VPS systems. Existing attacks are becoming more practical and more dangerous. In addition, new avenues of attacks to leak information are gradually being explored. Research on new areas of attack start by being creative and look to be more efficient, effective, and more general. On the other hand, many of the defenses have been tightly coupled with specific attacks presented in the research. More generic defense mechanisms are mainly being explored to target more popular attack mechanisms. All in all, further advancement in both VPS attacks and defense is likely to be propelled by benefits of better machine learning algorithms. With the improvements over the years, state of the art defense and attacks are increasingly practical on a much wider range of VPSes. It will be important for manufacturers to keep up with the latest developments to implement adequate countermeasures to protect the end users.